\documentclass[12pt]{iopart}

\usepackage{iopams} 
\usepackage{graphicx}
\usepackage[utf8]{inputenc}
\usepackage[english]{babel}
\usepackage{csquotes}
\usepackage{comment}
\usepackage[backend=bibtex, style=numeric, sorting=none]{biblatex}
\begin{document}

\title{Length measurement and stabilization of the diagonals of a square area laser gyroscope }

\author{N.~Beverini$^{1,2}$, G.~Carelli$^{1,2}$,A.~Di~Virgilio$^2$,
U.~Giacomelli$^{1,2}$, E.~Maccioni$^{1,2}$, F.~Stefani$^{1,2}$ and J~Belfi$^3$}

\address{$^1$University of Pisa, Pisa, Italy,}
\address{$^2$INFN Sezione di Pisa, Pisa, Italy}
\address{$^3$Altran, Milan, Italy}

\ead{giorgio.carelli@unipi.it}

\begin{abstract}
Large frame ring laser gyroscopes are top sensitivity inertial sensors
able to measure absolute angular rotation rate below prad/s in few seconds.
The GINGER project is aiming at directly measuring  
the Lense-Thirring effect with an $1\%$ precision
on an Earth based experiment. 
GINGER is based on an array of large frame ring laser gyroscopes. 
The mechanical design of this apparatus requires a micrometric precision 
in the construction and the geometry must be stabilized
in order to keep constant the scale factor of the instrument. 
The proposed control is based on square cavities, 
and relies on the length stabilization of the two diagonals, which must be
equal at micrometric level. GP2 is the prototype devoted 
to the scale factor control test. As a first step, the lengths of the
diagonals of the ring cavity have been measured through an interferometric
technique with a statistical accuracy of some tens of nanometers, 
and they have been locked to the wavelength 
of a reference optical standard. 
Continuous operation has been obtained over more than 12 hours, 
without loss of sensitivity. 
GP2 is located in a laboratory with standard temperature
stabilization, with residual fluctuations of the order of 1 $^{\circ}$C.
Besides the demonstration of the control effectiveness,
the analysis of the Sagnac frequency demonstrates that
relative small and low-cost ring lasers (around one meter of side) can
also achieve a sensitivity of the order of nrad/s in the range 
$0.01 - 10$~Hz in a standard environment, which is the target sensitivity
in many different applications, such as rotational seismology
and next generation gravitational waves detectors.	
\end{abstract}



\maketitle

	
	\maketitle
	
	\section{Introduction}
	
    Large frame Ring Laser Gyroscopes (RLG), which exploit
Sagnac effect, are the most sensitive devices for detecting absolute
angular motions in a huge range of frequency, extending
from kHz down to DC \cite{Schreiber2013, korth2016}. 
GINGER (Gyroscopes IN General Relativity) is a
project to build a large-frame 3-dimensional array of ring laser
gyroscopes to be mounted inside the underground
Gran Sasso Laboratory (LNGS). 
The Ginger final aim is to measure the Earth angular velocity
in the laboratory co-rotating  frame $\vec{\Omega}$, and 
to compare it with the Earth angular velocity as observed in the 
Cosmic inertial frame $\vec{\Omega}_\oplus$. 
General Relativity Theory (GR) foresees that in low field
approximation\cite{DiVirgilio2017} 

\begin{equation}
\vec{\Omega} = \vec{\Omega}_\oplus + \vec{\Omega}_{DS} + \vec{\Omega}_{LT} 
\label{GR}
\end{equation}

\noindent where $\Omega_{DS}$ (De Sitter or geo-electric effect) 
and $\Omega_{LT}$ (Lense-Thirring or geo-magnetic effect) 
take into account the GR effects  \cite{Bosi2011a} 
induced respectively by the Earth mass and the Earth angular momentum.
For this purpose it is necessary an instrumental accuracy of the order 
of $10^{-14}$~rad/s.
Such an accuracy offers also the possibility of measuring 
fundamental geodetic parameters related to the fluctuations 
of the Earth rotational velocity and of the Earth axis orientation, 
the so-called length of the day and polar motion. 
In the geodetic observatory of Wettzell (Germany)\cite{Schreiber2009}
the "Gross ring" (G) has already demonstrated a sensitivity of
$2\times 10^{-13}$~rad/s, not far from the GR test requirement.
Since G is based on a monolithic design, which cannot be extended 
to an array,  a new modular heterolithic design is required
for GINGER. To study this new idea and to validate the site, 
a square 3.6 m side ring-laser based on an heterolithic structure,
named GINGERINO [6],has been mounted in the underground
INFN Gran Sasso National Laboratories (LNGS). 
It showed that continuous data taking is feasible with high sensitivity 
and duty cycle over $95\%$ \cite{Belfi2017,Belfi:18}.

In general the response of a RLG to a rotation is given
by the  Sagnac frequency $f_s$:

\begin{equation}
f_s = \frac{4\ A}{\ P\ \lambda\ } \ \Omega\cos{\theta}
\label{sagnac}
\end{equation}

\noindent where $A$ is the area included by the optical path, $P$  the
perimeter length, $\lambda$  the laser wavelength, $\theta$ 
is the angle between the area versor of the RLG and the angular rotation
axis. For a RLG rigidly connected to the soil, $\Omega$ 
is the Earth rotation rate.
For a RLG lying on an horizontal plane,
 $\theta$ is the colatitude angle, while  for a RLG oriented
 with the axis parallel to Earth rotation axis $\theta$ is zero,
 and the Sagnac frequency has a maximum.
The GINGER design is based on a square cavity, 
and the control of the scale factor (the ratio area over perimeter)  
relies on the accurate measurement and control of the diagonal lengths. 
Indeed, the important feature of a square cavity, 
is that the four mirrors define three resonant cavities: 
the square one  and the two linear Fabry-P\'erot resonators 
along the diagonals.

At present the control scheme  \cite{Belfi2014,Santagata2015} is under test
by using a dedicated RLG named GP2. 
It foresees first to align the square cavity and 
then to control the geometry through  the length of the two diagonals.
In this way the heterolithic structure can guarantee the same performances,
in term of geometrical and thermal stability, of a monolithic one.
The  control procedure  will work in two steps. 
In the first step the two diagonals are carefully measured and compared.
Since the ring laser emission frequency is related 
to the ring perimeter length, it is possible to implement a procedure
to optimize the geometry of the ring optical path 
by acting on the corner mirrors,
as theoretically described in \cite{Santagata2015}.
It is possible to obtain a path which is as close as possible 
to a square by measuring the laser frequency at the same time with the diagonal lengths.
In the second step, during the GINGER operation, 
the optical path geometry will be  actively stabilized, by locking
the diagonal lengths to an external reference wavelength standard, 
also acting on the corner mirrors.
From an experimental point of view the diagonal lengths  
can be easily optically exploited, since  
they constitute two Fabry-P\'erot cavities, so that
standard metrological techniques can be used. 
In these conditions, according to our analysis,
a long term stability of the scale factor
of the order of  $10^{-12}$ can be obtained.
This is the needed stability for the $1\%$   test of the
Lense-Thirring effect, if the error in the positions of the mirrors 
with respect to the perfect geometry is lower than $1 \mu$m.
The GP2 prototype has been installed in a laboratory 
of the basement of the INFN Pisa building, with the purpose of
studying the experimental details of these procedures. 
In the following, after a brief description of the GP2 apparatus, we
will give an overview of the metrological techniques, 
and the first results 
on the scale factor control circuit are reported.

\section{Experimental Set-up}

\begin{figure}[htbp]
\begin{center}
\includegraphics[width=0.6\columnwidth]{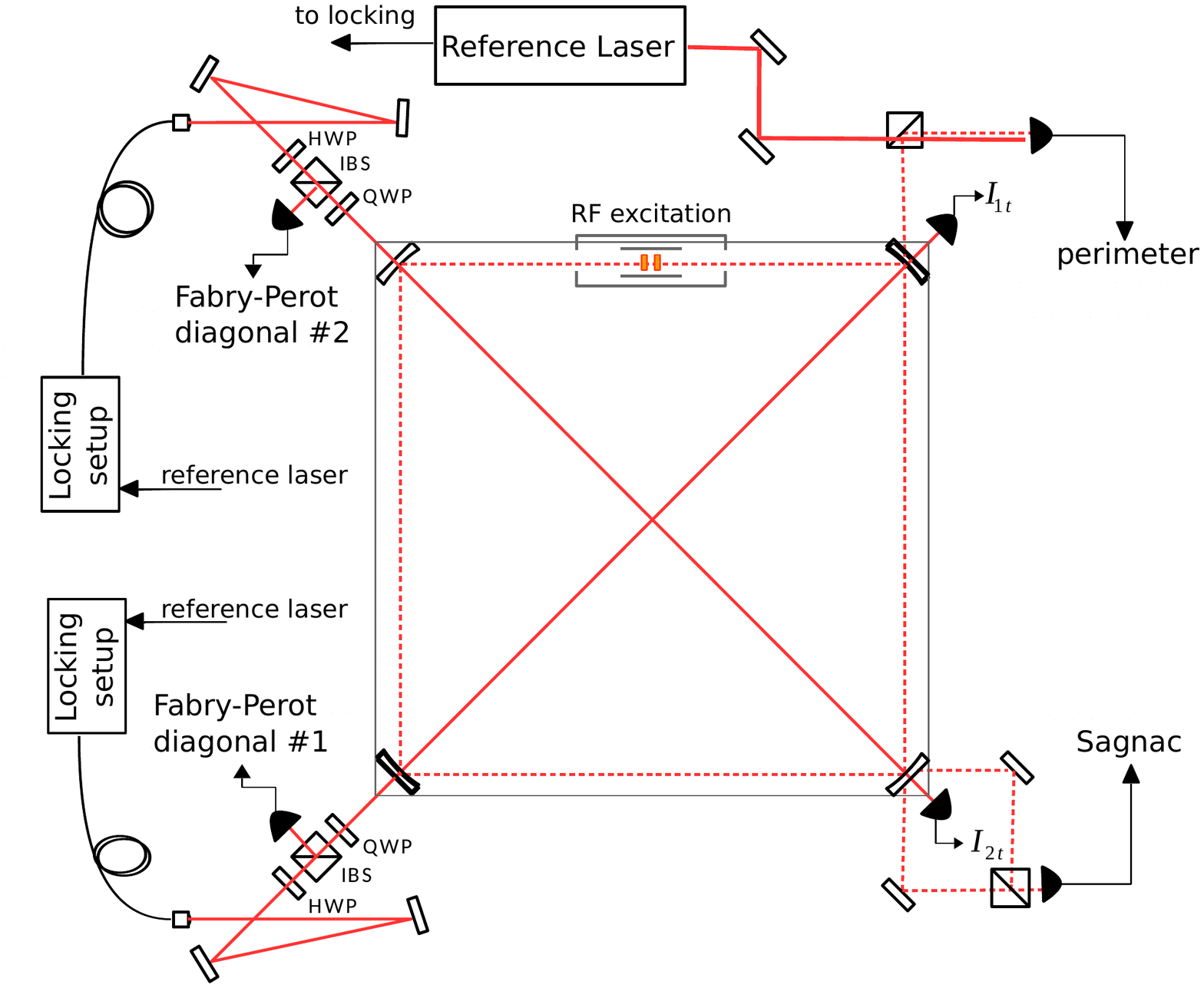}
\caption{Optical scheme of GP2 control setup}
\label{setup}
\end{center}
\end{figure}

GP2 is a square ring resonator, 1.6~m in side, 
mounted on a a granite table.
A pyrex capillary, 4 mm inner diameter, is inserted along one side,
where the gas discharge is excited. 
The gas is an He-Ne mixture, with a 1:1 ratio of the two isotopes
$^{20}$Ne and $^{22}$Ne \cite{Schreiber2009}. 
Typical partial pressures are of the order of 20 Pa and 500 Pa 
for Ne and He respectively. The scheme of the optical setup
is shown in Fig. \ref{setup}
The granite support is fixed to a concrete basement tilted 
following the local latitude, $43.4^{\circ}$. 
Then, the ring axis is almost parallel 
to the Earth rotation axis, so that the disturbances on the Sagnac signal
induced by local tilts are minimized.

GP2 has been built as a test bench to develop the  optics and the 
electronics to be implemented on GINGERINO and later on GINGER.
It is located inside a quite noisy environment, so that
very high sensitivity measurements are not feasible.
That's why we mounted on it high quality mirrors,
but not the top quality ones as used by GINGERINO. 
The mirrors have  2~m curvature radius and  $\sim1.5\times10^{-4}$ total
losses, estimated by a measured ring down time of $\sim 35$ $\mu$s.
 
The mechanical design of GP2 is quite different from that used 
in our previous prototype G-Pisa 
\cite{Belfi2012} and then in GINGERINO \cite{Belfi2017}.
Each corner mirror is mounted on a piezoelectric slide (PZT)
that can move along the diagonal direction
within an  $80\ \mu$m range. One of the corner is equipped with two more
piezo actuators in order to give a tri-axial Cartesian movement. 
As said above, the important feature of GP2 
is that the four mirrors define, besides the square ring resonator,
two Fabry-P\'erot along the diagonals.  
Each linear cavity is optically interrogated 
by injecting a light beam coming from the
same external stabilized laser source.
In order to avoid optical noise injection, this laser works at a  
slightly different frequency
from that of the ring laser emission .

The geometry stabilization procedure requires information both
on the ring perimeter and on the diagonal lengths.
The perimeter could be measured in two different ways. 
By using the self-beat note technique, or by measuring with an external reference.
In the first case the ring laser is driven
in a multimode regime, so that
different longitudinal modes generate on a fast photodiode
a beat-note at a frequency $n$-multiple
of the Free Spectral Range (FSR) of the ring cavity.
Then, the ring perimeter can be deduced as $c$/FSR .
This technique is very easy to be implemented, 
but multi-mode operation can produce
a decrease of sensitivity. This technique was successfully applied
to the G0 ring operating in Canterbury (NZ) \cite{Holdaway2012},
which had an excellent Sagnac signal  also in presence 
of multimode operation.  
It was however observed that the spectrum of the secondary modes 
were not stable. 
We tested  this technique on GP2, and we
observed a decay of the fringe contrast,
which in the worst case dropped from over $88\%$ to under $50\%$.
We then applied the  alternative technique,
by using an external laser as a reference. 
By combining one of the output
beams of the ring with the reference laser beam on a fast photodiode,
we obtain a signal at the difference of the frequencies.
This method requires a laser source with 
an high level of accuracy,
but it does not affect at all the quality of the Sagnac signal.
As a reference, we use a Helium-Neon laser stabilized on a saturated absorption transition of molecular Iodine; it was produced for us by the Institute of Laser Physics in Novosibirsk.
 \cite{Belfi2014}.
 
However, for the stabilization of the the scale factor, 
it's not sufficient the control of the perimeter,
and it's also needed to measure and to control accurately
the diagonal lengths \cite{Santagata2015}. 
For this purpose, a fraction of the reference laser radiation is injected, 
through a polarization maintaining single-mode fibers, 
into the Fabry-P\'erot resonators build by the diagonal opposite mirrors.
In a first step each cavity is locked through a Pound-Drever-Hall (PDH)
control circuit to the reference laser frequency, 
by acting on the mirrors by the (PZT).
In this way the cavity length is forced to be an integer multiple of the
laser wavelength.
In the second step, the absolute value of each diagonal lengths is
found, by measuring the frequency of a Voltage Controlled Oscillator (VCO),
while it is locked to a multiple $m$ of the Fabry-P\'erot free spectral range (FSR). 
The two values are acquired, and each cavity length $L$ is evaluated 
by using the relation $m\ c / 2L = m$FSR.
This technique was described in details in \cite{Belfi2014} and 
it was successfully tested on two 1.5 m long FP resonant cavities 
resting on an optical table. 

\begin{figure}[htbp]
\begin{center}
\includegraphics[width=0.8\columnwidth]{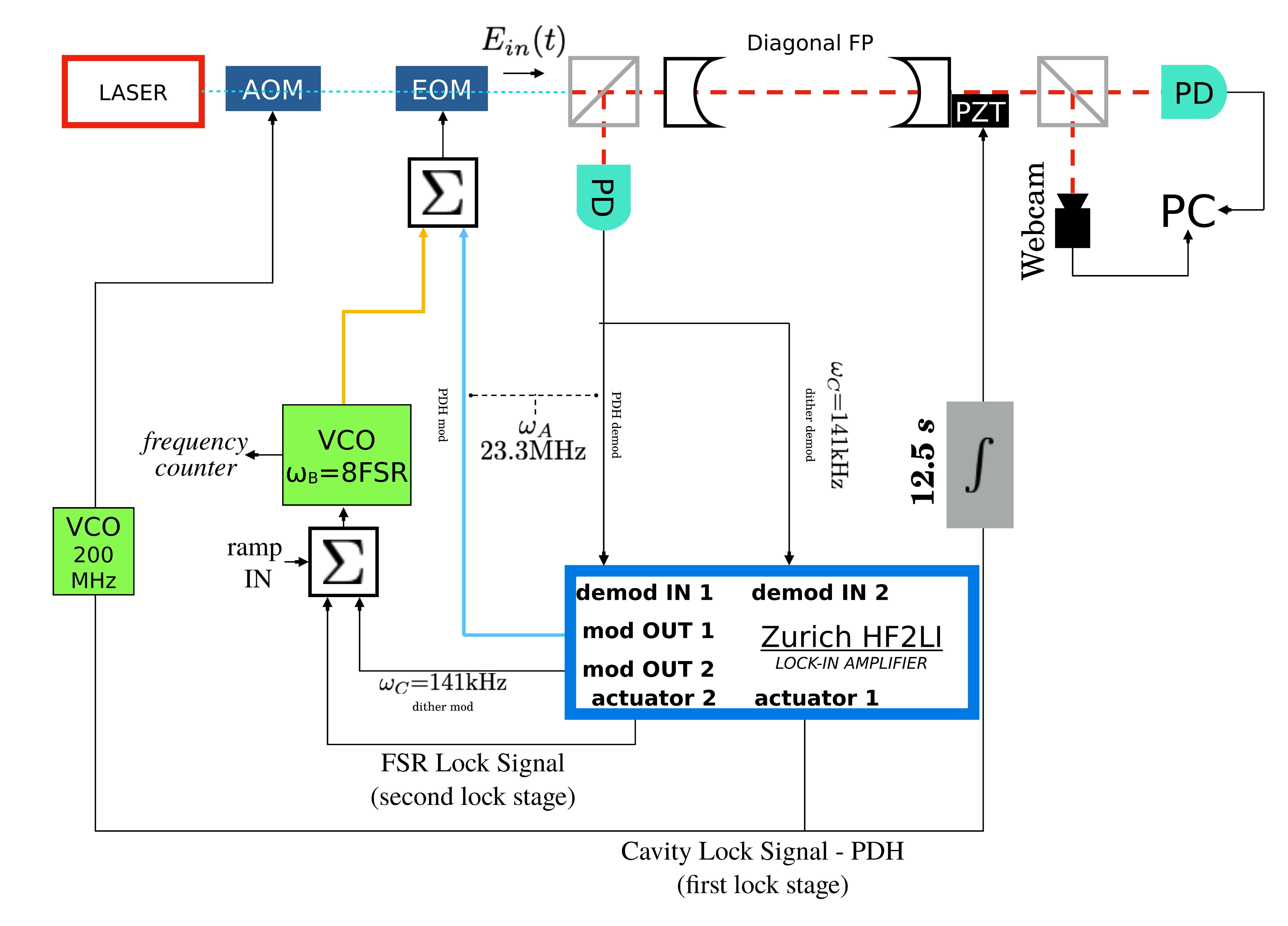}
\caption{Details of the electronic locking setups}
\label{Optical}
\end{center}
\end{figure}

The application to GP2 was not straightforward because 
the PZT actual bandwidth is reduced by the massive
vessel, whose mass is about 3~kg. 
The PZT effective bandwidth is no more than a few Hz,
while, in our laboratory condition, the cavities noise in the acoustic
band is larger, of the same order of the laser one.
Then, a simple PDH is not able to work correctly. 
That's why we separated the servo loop in two parts, and we implemented
different low frequency and high frequency controls. 
The low frequency component of the actuation signal is used, 
with an integration time of 12.5~s, to drive the PZT, 
while the high frequency component drives a VCO 
acting through an acousto-optics modulator (AOM)
on the frequency of the radiation in order to keep the
laser locked to the Fabry-P\'erot cavity resonance in the acoustic band. 
In short, at acoustic frequency, it is not the cavity 
that follows the laser,
but the noise that affects the cavity is added to the laser.
It is not a real correction, but it allows us to implement 
the slow stage of the lock loop. 
The addiction of this fast opto-electronic control improved greatly
the performance of the system and we obtained in this way 
the results discussed in the present paper.
The details of the  electronic circuits for measuring and  locking each diagonal are shown in Fig. \ref{Optical}. 
The injected laser field is modulated as:

\begin{equation}
       E_0 exp \{i(\omega _0 t  +
                         \alpha \sin(\omega _A t ) + 
                     \beta \sin [(\omega _B + \Delta \sin (\omega _C t )) t ])\}
\label{eqmod}
\end{equation}

\noindent where $\omega_0$ is the optical frequency of the carrier, 
$\omega _A \approx 23$~MHz is the PDH modulation frequency to
lock the carrier,
$\omega _B$ $= mFSR$ is the modulation frequency for the cavity 
dynamic resonance excitation (in our case FSR $\approx 66$~MHz and 
$m = 8$),
$\omega _C \approx 141$~kHz is the frequency of the dithering applied 
to $\omega _B$ for the down-convertion of the FSR resonance signal, 
and $\alpha$, $\beta$ and $\Delta$ 
are the respective modulation amplitudes. 

\noindent As shown in Fig. \ref{Optical}, the core of each control system is 
a Zurich Lock-in Amplifier HF2LI,
a digital lock-in that generates the two modulation frequencies $\omega _A$
and $\omega _C$,
demodulates the signals, and generates the two PID
(Proportional Integrative Derivative) correction signals
for the two locking steps.
The frequency of the VCO locked to the FSR is read by
a two channels frequency meter that was referred to a rubidium clock. 
The relative error of the diagonal length is proportional
to $1/m$, so we chose $m = 8$ 
as a compromise with the maximum allowed AOM frequency.

\section{Experimental results and data analysis}

\begin{figure}[htbp]
\begin{center}
\includegraphics[width=0.8\columnwidth]{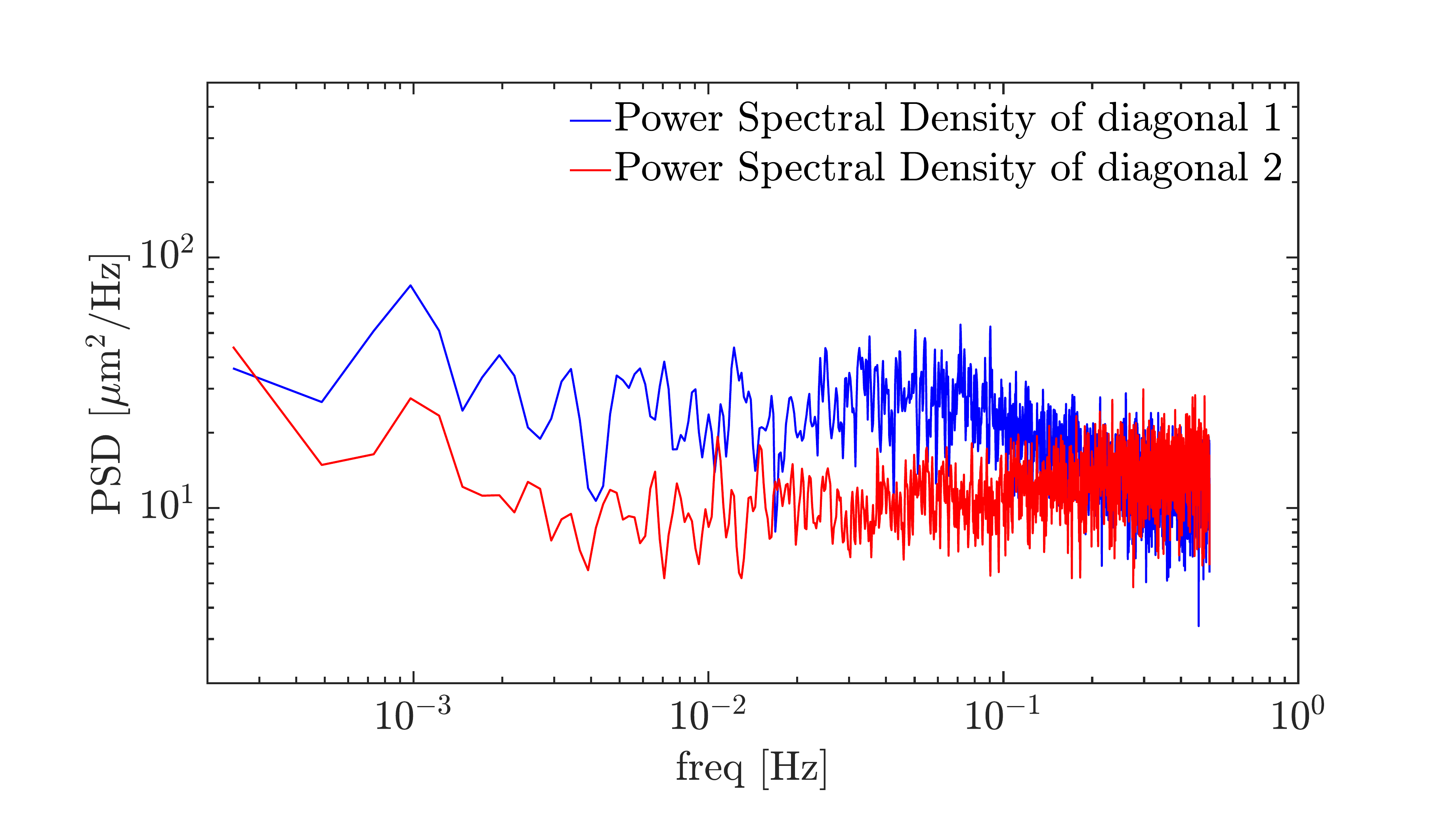}
\caption{Power Spectra of the diagonal cavities lengths}
\label{PSdiagonals}
\end{center}
\end{figure}

By using the optimized PID parameters generated by the Zurich Lock-in Amplifiers, 
we performed a simultaneous, about 12~hours long, measurement  
of both the diagonal lengths with a sampling rate of 1 data per second. 
The histogram of the results follows a Gaussian shape for both diagonals,
with a peak value of 2262.7699~mm and 2261.5515~mm 
and a Half Width Half Maximum, respectively, of 1.8~$\mu$m and 2.0~$\mu$m.   
The first evidence is that there is
a difference between the lengths of the two diagonals larger than expected. 
This difference is more than 1.2~mm, that is too large to be corrected
just acting on the PZT slides. 
The Gaussian profile of the measurements distribution
is demonstrated also by the flat trend of the power spectral density,
which shows a typical white noise behaviour, see Fig.\ref{PSdiagonals}. 
Then, we can also estimate the statistical accuracy of
both measurements as $\sim 19$~nm for the longest diagonal
and $\sim22$~nm for the shortest one. 
Both of these values are consistent with the requirements for GINGER. 
In addition, the absence of a flicker noise bias makes possible to increase
the accuracy increasing the measurement time.

\begin{figure}[htbp]
\begin{center}
\includegraphics[width=\columnwidth]{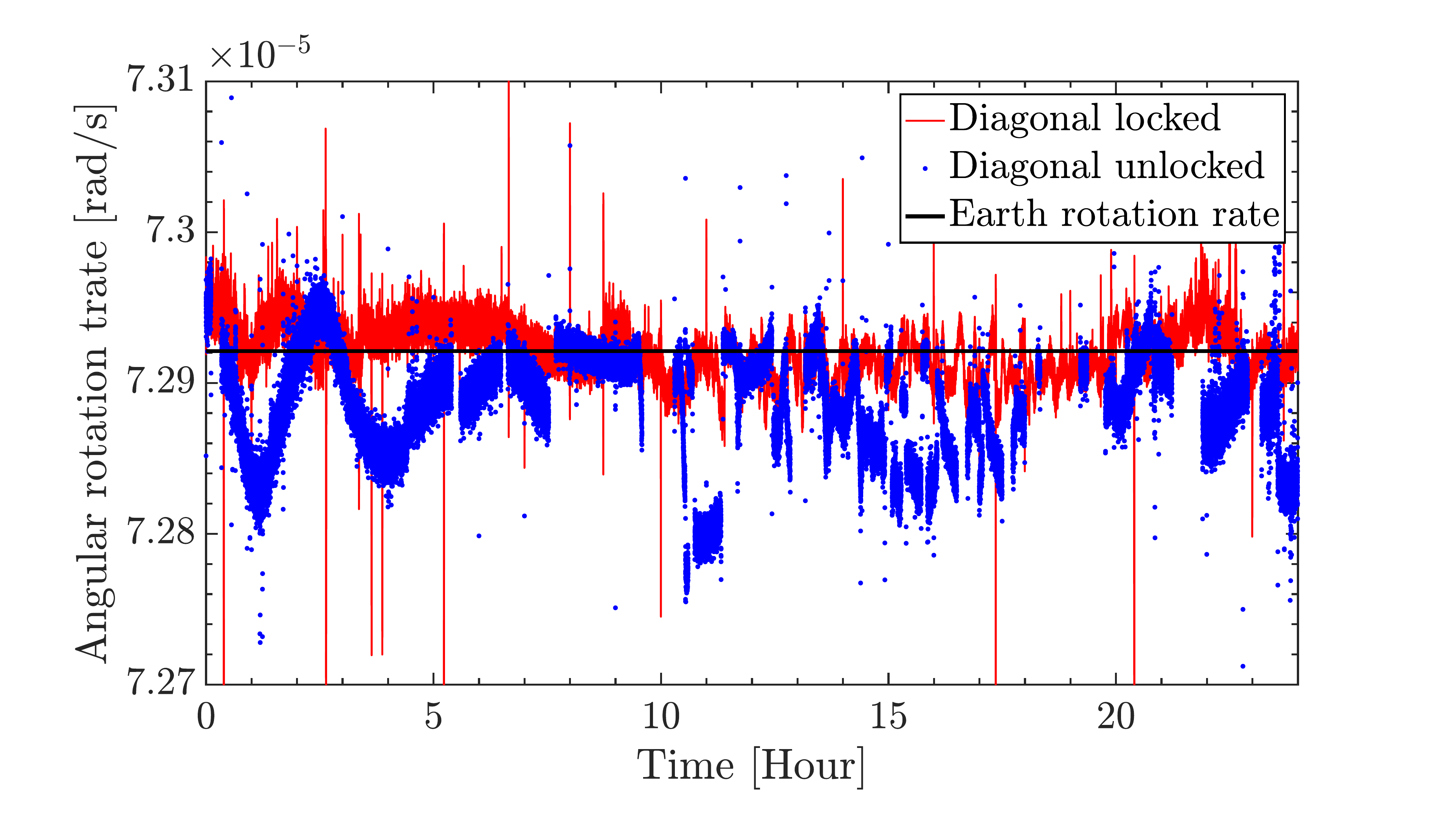}
\caption{Rotation rate reconstructed for the two runs: 
locked (red) and unlocked (blue). The black line shows the earth rotation rate.
The duty cycle in the locked mode is over $99\%$, 
while in the unlocked one drops down to $78\%$ Split modes and mode jumps
have been eliminated by using the information of the fringe contrast. 
The large peaks present in the locked data are disturbances due to
human activity around the laboratory, which is located in the university area.}
\label{LUn}
\end{center}
\end{figure}

We have tested the effectiveness of the technique of locking the diagonal
length by measuring  GP2 Sagnac
signal over one day both with diagonals locked and unlocked. The data were
processed following the analysis scheme developed for GINGERINO
\cite{Belfi:18}.
The Sagnac frequency, the intensities of the two counter-propagating beams
and the monitor of the discharge fluorescence were recorded at 5 ksample/s
acquisition rate. In the analysis, the portions of data affected 
by split mode or mode jumps were removed \cite{Belfi:18}. 
Fig. \ref{LUn} compares the reconstructed angular
rotation rate in locked (red curve) and 
in free running condition (blue curve). 
In locked operation  the duty cycle was over $99\%$,
while in free running condition  it dropped down to $78\%$. 
It is as well evident that the free running data show larger perturbations. 

GP2 is oriented at the maximum Sagnac signal; then it should measure the
modulus of the Earth rotation rate. 
From data taken during locked operation, we can evaluate an averaged value 
of the rotation rate 
$\Omega_m = 7.2919\ (3)\times10^{-5}$~rad/s, which is consistent with the value 
of Earth rotation rate (Fig. 4, black horizontal line)  $\Omega_\oplus =  7.29211509 \times 10^{-5}$~rad/s.

GP2 is affected by large perturbations, being located in the basement 
of a building of the Physics Department of the University of Pisa in a
laboratory where the temperature is stabilized at $\pm 0.5$~K. 
The long term fluctuation of the average value of the reconstructed angular 
rotation rate in free running operation is however of the order of 
$3 \times 10^{-4}$, that is
more than one order magnitude larger then expected  by simply considering 
the  effect on the scale factor of the thermal expansion. We can explain 
this large deviation as an effect of the back-scattering of the laser
radiation
on the optical cavity mirrors, which is very sensitive to the geometry,
as discussed in  \cite{Belfi:18,Belfi2010}.

\begin{figure}[htbp]
\begin{center}
\includegraphics[width=\columnwidth]{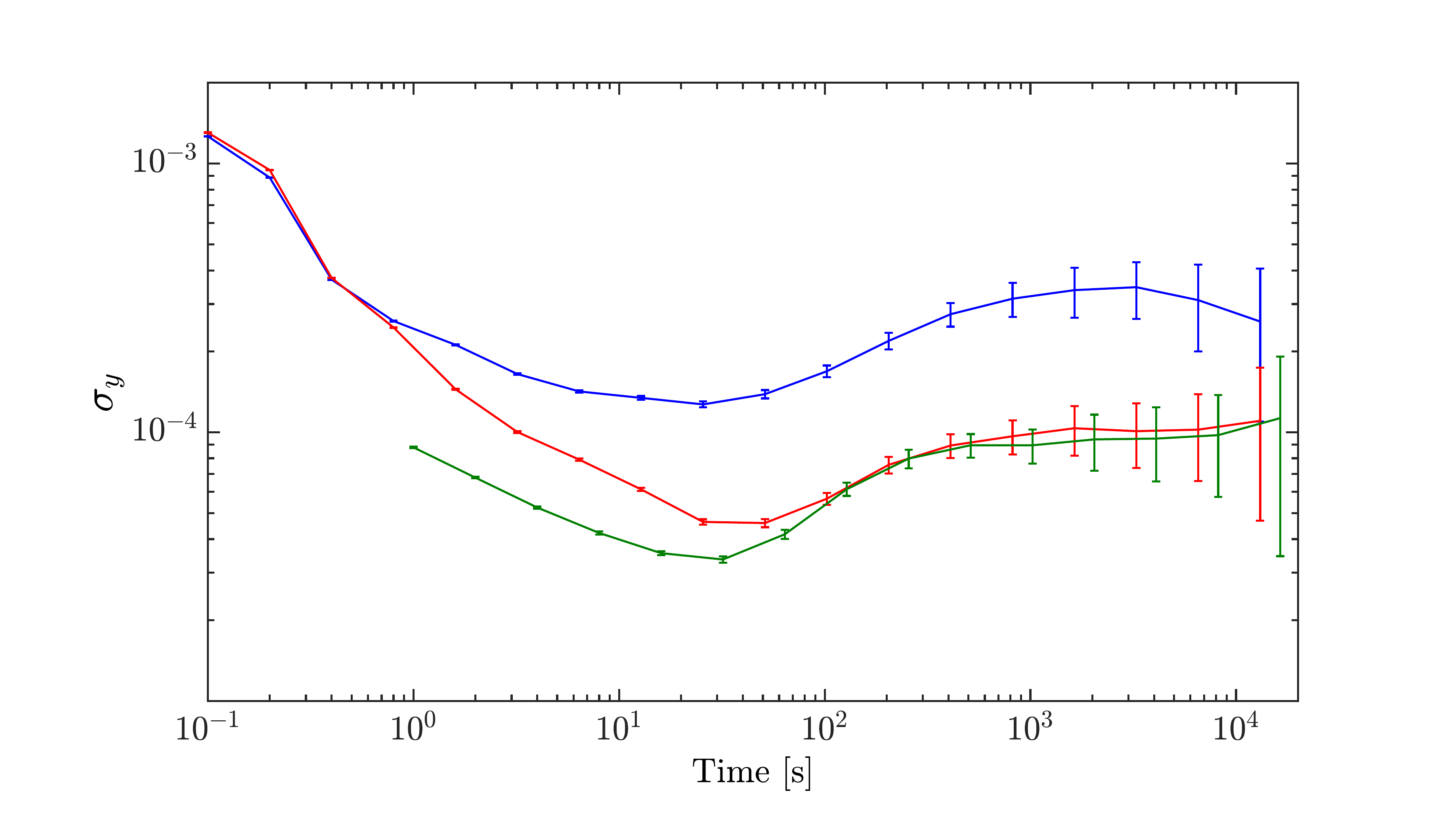}
\caption{Allan deviation of Sagnac signal in three different cases.
1: diagonals unlocked at 10 Hz sampling, blue, 
2: diagonals locked at 10~Hz sampling, red, and 
3: diagonals locked at 1~Hz sampling, green. 
The best sensitivity is $2$ nrad/s obtained in 30s measurement.}
\label{Allan}
\end{center}
\end{figure}

 Fig. \ref{Allan} shows the Allan deviation in the two cases.
 A sensitivity better than of $10$ nrad/s is obtained at 1~second of measurement, 
 with a minimum of $\sim 2$~nrad/s at $30$~s of integration time.

\section{Discussion and Conclusions}

We showed a scale factor control technique for a large frame RLG system,
which is based on the active control of the lengths of the diagonals  
of the square cavity, and on their accurate measurement . 
The technique to measure the diagonal lengths has been developed 
and tested on the GP2 prototype. The obtained accuracy
is compliant with the requirements for GINGER ($1\ \mu$m error in the
difference between the two values), which is aiming 
at the Lense-Thirring test at $1\%$ precision.  
Two test runs with diagonals locked and unlocked have been performed.
The analysis has shown a duty cycle around $99\%$,
 while the active locking is on and around $78\%$ in the uncontrolled case.
In term of sensitivity the short time response is similar in the two cases, while the long term one is a factor 3 better in the controlled case.
The test has shown as well the necessity to expand the bandwidth 
of the control loop. 
For this purpose it will be necessary to reduce the load of the 
actuators; in the present scheme the PZT actuators 
move the whole mirror holder. 
Considering that GP2 is done utilizing granite and steel, 
with our laboratory thermal conditions the long term 
stability of the scale factor should be around 1 part $10^{5}$. 
The observed one is a factor $10$ worse, further studies and tests are 
necessary in order to investigate whether the long term response 
could be improved. An important improvement would be obtained by a better
equalization of the two diagonal lengths, which is unfeasible
with the present apparatus.
Moreover, GP2 is located in a quite noisy environment 
and its mirrors are not top quality ones.
In the future the test will be repeated by installing better mirrors,
by acting on both mirrors for each cavity, 
in order to reduce changes on the back-scattered light,
and by improving the shielding of the the RLG from external disturbances
such as air flow, etc. 
  
Anyway, this test shows that a middle size RLG can
work continuously in a standard laboratory with a sensitivity 
of the order of a few~nrad/s in the range 0.01-10~Hz. 
A range that can be extended at lower frequency by an effective geometry control. 
This result also demonstrates that relative small, 
with an area of the order of 1 $m^{2}$, RLGs have a potential  utility 
as very sensitive sensors of ground tilting. 
We note that Sagnac effect is sensitive to the rotation 
relative to an inertial frame and therefore is not affected
by Newtonian noise while, on the contrary, balance tiltmeter are 
sensitive to rotation relative to the local $\vec{g}$ direction.
This kind of properties is rather timely, for application in future
generation gravitational waves research and for seismological study in general.


	

\printbibliography


\end{document}